# Role of Zn and Nb in the Giant Piezoelectric Response of $PbZn_{1/3}Nb_{2/3}O_3$

A. Al-Zein[†,‡,§,∥], H. Dammak[‡], Ph. Papet[†], O. Mathon[§], B. Hehlen[∥], C. Levelut[∥], J. Haines[†], J. Rouquette[†*]

[†]Institut Charles Gerhardt UMR CNRS 5253 Equipe $C_2M$, Université Montpellier II, Place Eugène Bataillon, cc1504, 34095 Montpellier cedex 5, France.

[‡]Laboratoire Structures, Propriétés et Modélisation des Solides, UMR 8580 CNRS-École Centrale Paris, Grande Voie des Vignes, F-92295 Châtenay-Malabry Cedex, France

[§]European Synchrotron Radiation Facility, 6 rue Jules Horowitz, B.P. 220 38043 Grenoble cedex 09, France.

[∥]Laboratoire Charles Coulomb CNRS/UMR5221, cc 69, 34095 Montpellier cedex, France.

*Supporting Information Placeholder*

**ABSTRACT:** Relaxor ferroelectrics perovskite are highly polarizable and can exhibit giant coupling between elastic strain and an applied electric field. Here, we report an "in situ" extended X-ray absorption fine structure (EXAFS) study of a $PbZn_{1/3}Nb_{2/3}O_3$ (PZN) single crystal as a function of the electric field. We show that the strong dipoles in the $NbO_6$ octahedra bonds are aligned along the 4 <011> directions close to the orientation of the electric field, while a small reversible polar shift occurs for Zn in the direction of the electric field, i.e. positive or negative. This reversible Zn-O polar shift is proposed to play an important role in both the "easy" switching of the ferroelectric polarisation and of the giant piezoelectric effect in PZN.

## Introduction

The growth of $PbZn_{1/3}Nb_{2/3}O_3$ (PZN) and PZN-based perovskite single crystals has allowed a significant advance in high piezoelectric efficiency to be obtained,[1] which can find a large number of technological applications (actuators, sensors, microelectromechanical systems)[2]. These materials exhibit ultra-high piezoelectric coefficients $d_{33}$ from 1100 to 2500 pC/N in PZN and PZN-8%PT (a PZN solid solution with 8% of $PbTiO_3$) respectively for <001> oriented crystals[3,4], which can be compared with the 220 pC/N in the well-known lead zirconate titanate ceramics (PZT and 550 pC/N for doped PZT)[5]. The origin of the large dielectric susceptibility and piezoelectric properties in these highly polarizable ferroelectric perovskites of general formula $ABO_3$ is due to the cooperative displacement of the B cations, i.e. the Zn/Nb atoms in our case, from the center of symmetry of their octahedral sites. These spontaneous displacements create local dipoles that are aligned parallel to one another in order to give a macroscopic dipole moment by the application of an electric field $E$. In the case of $PbMg_{1/3}Nb_{2/3}O_3$ and PZN single crystals, the average $ABO_3$ pseudocubic perovskite structure is perturbed by translational disorder on an intermediate length scale [6,7]. It is known that this nanoscopic inhomogeneity is responsible for relaxor properties of these materials, such as smearing and frequency dependence of the dielectric anomaly associated with the ferroelectric ordering [8] and for phenomena implying the presence of polar nanoregions (PNR) persisting hundreds of degrees above the temperature of the dielectric permittivity maximum [9,10]. While focusing on the PZN-PT solid solution, one can remark the large difference in piezoelectric properties of the two end-members $PbTiO_3$ ($d_{33}$= 35 pC/N) [11] and $PbZn_{1/3}Nb_{2/3}O_3$ ($d_{33}$= 1100 pC/N) [1], which have tetragonal ($P4mm$ space group) [12] and pseudocubic structures (with small rhombohedral distortion) [13] respectively. As long range order in ferroelectric relaxor is not properly defined, it appears more appropriate to differentiate these two crystals based on their individual atomic arrangements[14-18]. This is why we decided to perform an "in situ" extended X-ray absorption fine structure study of a $PbZn_{1/3}Nb_{2/3}O_3$ single crystal as a function of the applied electric field to clarify the individual role of the displacements of the Zn and Nb atoms. In the PZN-PT solid solution, the role of Pb atoms is of particular importance as it hybridizes with oxygen states leading to a large strain linked to the piezoelectric effect [19]. Additionally, if lead atoms are replaced by barium atoms, i.e. $BaZn_{1/3}Nb_{2/3}O_3$, 1:2 ordered structures (1 Zn/ 2 Nb) with a trigonal distortion space group $P\bar{3}m1$ are observed, which is a clear indication of the specific role of lead in the disordering on B-site positions[20]. Hence, the clear difference in the $ABO_3$ perovskite between $PbTiO_3$ and $PbZn_{1/3}Nb_{2/3}O_3$ is based on the B-site elements. In $PbTiO_3$, small $Ti^{4+}$ ions (r($Ti^{4+}$) = 0.605 Å) [21] exhibit pronounced off-center shifts (0.3 Å) [22] in their oxygen octahedra, giving rise to the strong spontaneous polarization ($P_S$ = 75 $\mu C/cm^2$) [23] along the [001] tetragonal direction. Looking at the unoccupied or fully occupied d orbitals, ($d^0$ state of the $Nb^{5+}$ and the $d^{10}$ state of the $Zn^{2+}$ ions), it is possible to have an insight into the local shifts on the same site, i.e. the center of the oxygen octahedron. $Zn^{2+}$ and $Nb^{5+}$ have ionic radii of

0.74 and 0.64 Å [21], respectively. One can therefore expect a much more pronounced off-center shift of smaller Nb atoms in their oxygen octahedral cage in comparison to the larger Zn atoms. This scenario was confirmed based on X-ray absorption spectroscopy (XAS) experiments; the Nb off-center displacement is experimentally found to be about 0.1 Å, while Zn exhibits no shift [24, 25]. Based on the information on B-site elements described above, one can predict that the large Ti-off center shift in PbTiO$_3$ will require a large coercive field (6.75 kV/cm) [23] to flip the spontaneous polarization, which can be linked with the relatively low piezoelectric coefficient of lead titanate. In contrast, PZN exhibits a smaller average off-center shift, which can explain the low coercive field (2.7 kV/cm) needed to reverse the smaller $P_S$ ($P_S$ = 25 µC/cm$^2$) [1]. Hence, the B-site polar character in PZN arises predominantly from the Nb-atoms. However, the mechanism underlying the ultrahigh performance of PZN and more particularly the electric field dependence of the Zn and Nb atoms is the key issue to give the appropriate atomic scale picture of this functional material under "in-situ" conditions corresponding to those experiments in technological applications. Therefore, XAS appears as the technique of choice to clarify the individual role of the displacements of the Zn and Nb atoms as a function of the applied electric field. Here we report an "in situ" EXAFS study on PZN single crystal poled along the [001] direction, i.e. the direction of the giant piezoelectric effect.

### Experimental Section

EXAFS measurements were performed on PZN plates metalized with gold (typically 2.5 mm × 2.5 mm × 200 µm), with large surface perpendicular to the [001] direction, at beamline BM29 of the European Synchrotron Radiation Facility. Direct voltage was applied between the electrodes along the [001] direction during the measurements. A (311) Si monochromator with a resolution of 0.5 eV was used. Because of the large absorption coefficient of lead, EXAFS data were collected in fluorescence mode at both the Zn $K$ edge (9659 eV) and the Nb $K$ edge (18986 eV). The fluorescence mode appears to be an appropriate geometry to investigate the electric field dependence of the crystal, i.e. 45° scattering geometry between the X-ray incident direction electric field vector $\vec{e}$ and the applied electric field $\vec{E}$ linked to the polar shift. The data have been corrected to the self-absorption approximation, which affects the amplitude of EXAFS spectra measured in fluorescence mode. To avoid any correction errors, we checked that the EXAFS spectrum of a pellet of finely ground powder of PZN and boron nitride measured in transmission geometry gives similar results as the fluorescence spectrum of our unpoled PZN single crystal. It is worth noting that no changes are expected in the XANES and the pre-edge regions in this system especially the $p$-$d$ peak, which arises from mixing of unoccupied $p$ and $d$ states of the B atom[26], because the $d$ orbitals of the Zn atom are fully occupied ($d^{10}$ state) and Nb $K$ hole width is large ($\gamma K$(Nb) ≈ 4.14 eV). The EXAFS data analysis was carried out with the IFEFFIT package [27]. The contribution of interest to the total EXAFS signal is given by the single scattering between the photoabsorber, i.e. the B-atom (Zn or Nb), and the first six oxygen neighbours over the whole electric field range, i.e. 0 ≤ $E$ ≤ 8 kV/cm. The time to measure one EXAFS dataset as a function of $E$ was about 75 minutes. At least two spectra per electric field value were needed to give reproducible data. Attempts were carried out to measure EXAFS spectra during the first polarization cycle of the PZN crystal. However, kinetic effects during polarization are well-known in ferroelectric PZN crystals [28], i.e. the electric field necessary to pole the sample strongly decreases with time. This means that the poling state of an unpoled sample under electric field can vary during acquisition and the set up used in this study does not allow us to accurately describe the EXAFS spectra of the first polarization cycle in PZN. Thus, we will only present measurements on the unpoled state and the poled state under electric field.

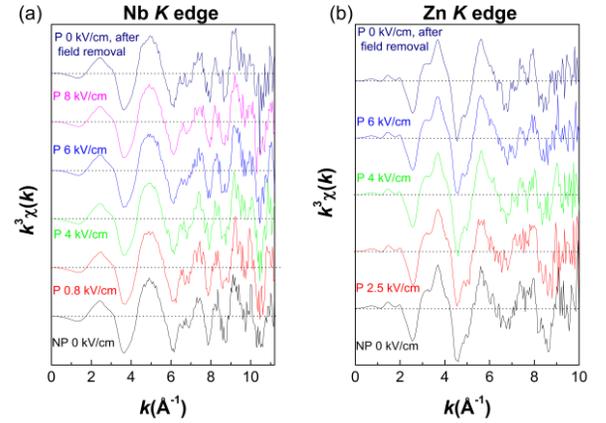

**Figure 1.** EXAFS spectra of PbZn$_{1/3}$Nb$_{2/3}$O$_3$ crystal as a function of electric field. (a) Nb $K$ edge spectra, (b) Zn $K$ edge spectra. NP and P accounts for non poled and poled state.

### Results

High quality Nb $K$ edge and Zn $K$ edge EXAFS data of PZN were obtained in the 0 ≤ $E$ ≤ 8 kV/cm electric field range, Figure 1. $k^3\chi(k)$ spectra are shown here to better take into account the contribution of the Nb(Zn)-O first coordination shell, which is linked to the fast damping of the scattering from the O-atoms with increasing $k$. In the case of the Nb-EXAFS data, Figure 1a, one can observe that the electric field has no drastic effect on the positions of the EXAFS fine structure, which indicates that the Nb local environment is not strongly affected during poling. However, the oscillations exhibit stronger amplitudes at high $k$ between 8 and 11.3 Å$^{-1}$ for the poled samples. This can clearly be associated with the EXAFS Debye-Waller factor: exp ($2\sigma^2k^2$)), which accounts for the relative mean square displacement linked to the Nb-O distances and is therefore an indication of order/disorder.

Let us now discuss the Zn-EXAFS data, Figure 1b. As in the spectra at the Nb $K$ edge, one can observe that the electric field has no strong effect on the positions of the EXAFS fine structure, which is consistent with a similar local environment of Zn atoms during poling. A careful inspection of the EXAFS oscillations shows that the unpoled crystal and the poled crystal at 2.5 kV/cm exhibit higher amplitudes between 6 and 10 Å$^{-1}$ compared with the poled samples at 4 and 6 kV/cm respectively. However, when the electric field is removed (poled state 0 kV/cm) the pronounced high $k$ oscillations reappear. Again, the main effect of the electric field on the Zn *K* edge EXAFS signal will be based on the order-disorder around the Zn-atoms and its associated $\sigma^2$, but in a different manner when compared with the Nb-data.

### Nb-O ordered model

Typical Nb-O first coordination shell in the *k* space for different electric field values are shown on Figure 2. Fitting of the EXAFS signal was done in *k* space isolating the Nb-O first coordination shell between 1.1 Å and 2.1 Å and back transforming in *k* space ($2 \leq k \leq 11.3$ Å$^{-1}$) using a standard procedure.

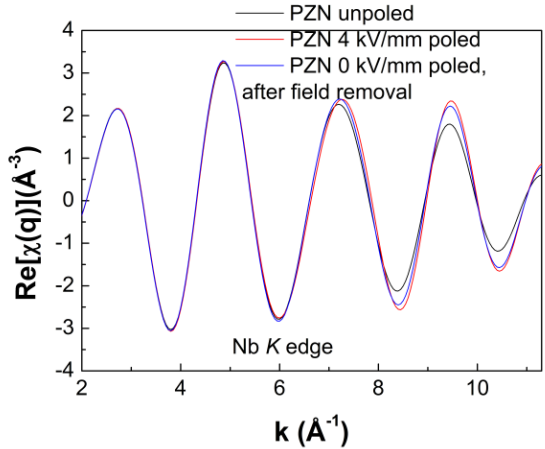

**Figure 2**. Inverse Fourier transform of the Nb-O atom pair distribution function in the $k^3$-weighted space calculation obtained for an unpoled sample (black), a sample poled at 4 kV/cm (red) and a poled sample (blue) after the electric field was removed (0 kV/cm).

All the different structural models were tested for the Nb-O first coordination shells, i.e. orthorhombic (tetragonal), rhombohedral and cubic local symmetries. We have to keep in mind that ferroelectric material can exhibit either no B-site off-center, with the corresponding cubic local structure, or a polar distortion, rhombohedral or tetragonal; this latter tetragonal distortion can only be orthorhombic (4-2) with a corresponding dipolar moment along the <110> direction, Figure 3, as the Jahn-Teller tetragonal distortion presents no polar shift.

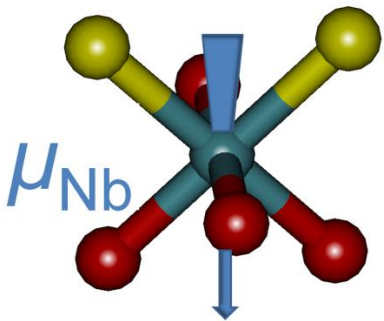

**Figure 3**. Orthorhombic 4 short (red, $R_1$ = 1.94 (1) Å) – 2 long (yellow, $R_2$ = 2.15 (1) Å) distribution of Nb-O bond distances along with <110> Nb-off center $\mu_{Nb}$.

Based on Figure 4 which shows these different solutions for unpoled PZN, we can obviously show that Nb-O can only be described by an orthorhombic distribution of bond distances (4-2).

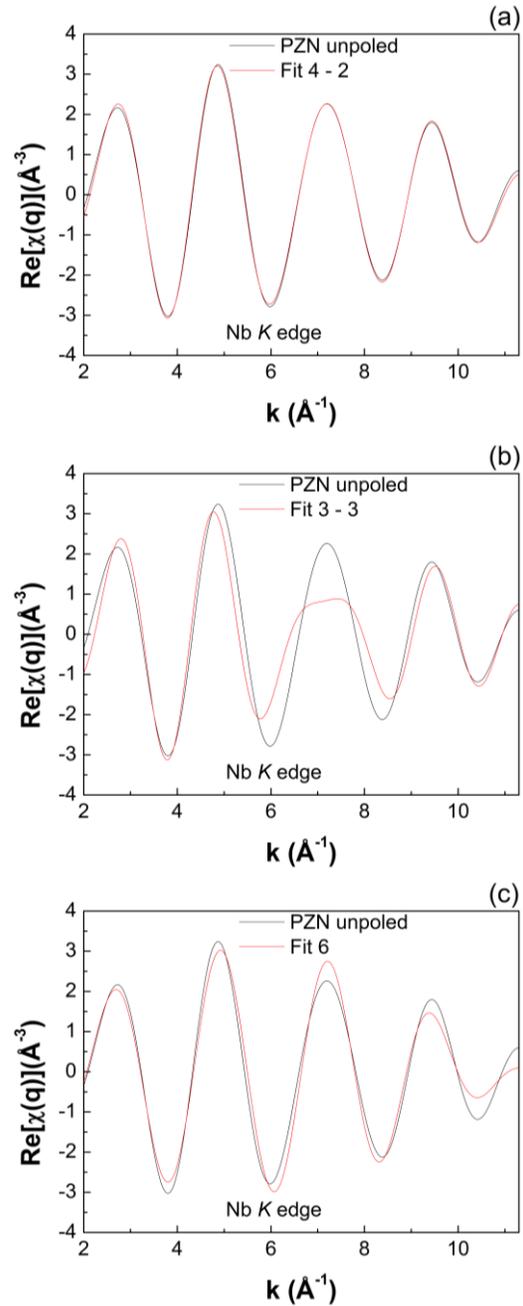

**Figure 4**. Inverse Fourier transform EXAFS signal corresponding to the Nb-O atom pair distribution function along with the best fit calculation obtained for the unpoled crystal using: (a) Orthorhombic ($R_1$ = 1.94 (1) Å), $R_2$ = 2.15 (1) Å) distribution of Nb-O bond distances. (*R-factor* = 0.5%), (b) rhombohedral 3 – 3 ($R_1$ = 1.97 (4) Å, $R_2$ = 2.19 (4) Å) distribution of Nb-O bond distances (*R-factor* = 14.5 %), (c) cubic - 6 (R = 1.92 (1) Å) distribution of Nb-O bond distances (*R-factor* = 3.6 %). This orthorhombic distribution of bond distances was therefore used, characterized by the 3 following refined structural parameters: $R_1$ (four equivalent short distances, $R_1$ = 1.94 (1) Å) and $R_2$ (two equivalent long bond distances $R_2$ =

2.15 (1) Å), $\sigma^2$(Nb-O) (relative mean square displacement or EXAFS Debye-Waller factor (DWF): exp ($2\sigma^2k^2$)) which was constrained to be the same for $R_1$ and $R_2$, and the number of neighbours, which was fixed to six. $\sigma^2$(Nb-O) contains two contributions: a thermal ($\sigma^2_{dyn.}$) and a structural ($\sigma^2_{stat.}$) contribution linked to the local vibrational dynamics and the local structural distortions, respectively. In this study, as the thermal contribution is constant, changes in $\sigma^2$(Nb-O) correspond to a structural contribution, linked with the orthorhombic distortion constrained by $R_1$ and $R_2$. This orthorhombic Nb-O environment, Figure 3, was already found to be a common feature in all the PbB'NbO$_3$ perovskite compounds (B'=Zn, Mg, Sc, In) and is independent of the macroscopic symmetry (cubic, rhombohedral, orthorhombic) [24, 25]. Such a local arrangement is also very close to what was observed for the Ti/Zr atoms in the lead zirconate titanate solid solution with a 5-1 distribution of the bond distances (in PbZr$_{0.20}$Ti$_{0.80}$O$_3$: [4×1.988 Å +1×1.907 Å] + 1×2.242 Å) [29], which is also in agreement with our data for PbZr$_{0.52}$Ti$_{0.48}$O$_3$ [30]. Figure 2 shows the backtransformed EXAFS signal corresponding to the Nb-O atom pair distribution function, i.e. backward Fourier transform, in the $k^3$-weighted space calculation obtained for an unpoled sample, a sample poled at 4 kV/cm and a poled sample where the electric field has been removed (0 kV/cm). As already observed in the Nb K edge spectra[21], one can clearly observe that the oscillations of the Nb-O atom pair distribution function between 8 and 11.3 Å$^{-1}$ exhibit a higher amplitude (and identical positions) for the poled sample at 4 kV/cm and 0 kV/cm respectively, compared to the unpoled state, which is a clear indication of a change in the order in the Nb-O local environment. Figure 5a and 5b show the inverse Fourier transformed EXAFS signal along with the best fit calculation obtained for the poled sample at 4 kV/cm and 0 kV/cm after field removal respectively. The experimental data are correctly reproduced over the electric field range investigated.

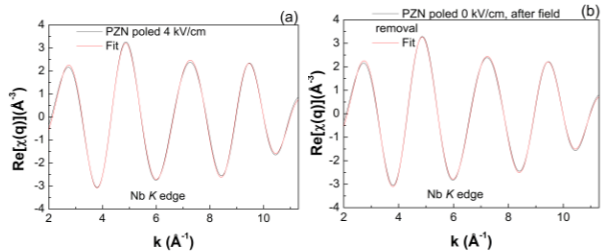

**Figure 5**. Inverse Fourier transform EXAFS signal corresponding to the Nb-O atom pair distribution function along with the best fit calculation obtained for (a) the poled sample at 4 kV/cm and (b) the poled sample after removal of the electric field (0 kV/cm).

### Zn-O disordered model

Let us now discuss the Zn-EXAFS data. Here, we refined the k space spectra isolating the Zn-O first coordination shell between 1.2 Å and 2 Å and the back transforming in k space ($2 \leq k \leq 10$ Å$^{-1}$) using a standard procedure. Over the entire electric field range, a symmetric distribution of bond distances was used for Zn-O as no "clear distortion" was observed up to the highest electric field value as explained below. Additionally as already mentioned in the introduction, the large Zn atoms are supposed to exhibit no polar shift and this symmetric distribution of bond distances is clearly appropriate as in the case of the Zr atoms in PZT [31]. We attempted to use rhombohedral and tetragonal models to fit our data, i.e. with 2 distances $R_1$ and $R_2$. Below 4 kV/cm $R_1=R_2$, the data are consistent with the proposed symmetric model, whereas a distortion could be observed for E $\geq$ 4 kV/cm although the estimated standard deviation was found to be of the order of the proposed distortion. Additionally, as the momentum transfer is limited to 10 Å$^{-1}$ in our data, both rhombohedral and tetragonal structural models gave similar agreement factors. Therefore, we decided to keep the symmetric distribution of bond distances for the entire electric field range as all the data are analyzed with the same structural model. In such a disordered model as the thermal contribution is constant, the increase in $\sigma^2$(Zn-O) thus corresponds to a polar structural contribution μZn, i.e. ΔR between $R_1$ and $R_2$ is therefore contained in the DWF. Additionally such a disordered model only uses two refined parameters $R$ and $\sigma^2$(Zn-O) instead of at least three for an ordered model ($R_1$, $R_2$, $\sigma^2$(Zn-O)), which is particularly interesting using the restrained momentum transfer range obtained in our data. In the entire electric field range we used the two refined following structural parameters: $R$ (average bond distance, $R$ = 2.066 (5) Å), $\sigma^2$(Zn-O), and the number of neighbours which was fixed to six. The backtransformed EXAFS signal for an unpoled sample, a sample poled at 6 kV/cm and a poled sample after removal of the electric field (0 kV/cm) is shown in Figure 6.

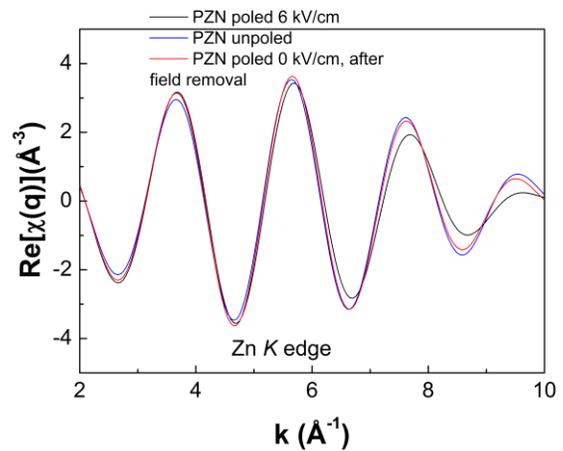

**Figure 6**. Backward Fourier transform in the $k^3$-weighted space calculation obtained for an unpoled sample (blue), a sample poled at 6 kV/cm (black) and a poled sample (red) after removal of the electric field (0 kV/cm).

As already observed in Zn-EXAFS data, Figure 1b, it is possible to note that the oscillations of the Zn-O atom pair distribution function exhibit a stronger amplitude (and identical positions) between 6 and 10 Å$^{-1}$ for the unpoled and poled (0 kV/cm) sample, compared to the poled (6kV/cm) state, which shows a change in the order in the Zn-environment. Figures 7a and 7b show the backtransformed EXAFS signal along with the best fit calculation obtained for the poled state at 0 and 6 kV/cm respectively. The experimental data are correctly reproduced over the electric field range investigated.

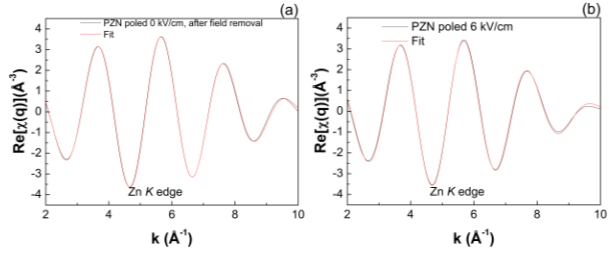

**Figure 7**. Inverse Fourier transform EXAFS signal corresponding to the Zn-O atom pair distribution function along with the best fit calculation obtained for (a) the poled crystal (0 kV/cm) after field removal and (b) the poled state (6 kV/cm).

## Discussion

Figure 8a shows the electric field dependence of the experimental $\sigma^2$ for the Nb-O atom using the orthorhombic ordered model. A noticeable decrease of $\sigma^2$(Nb-O) can be observed between the unpoled and the poled state.

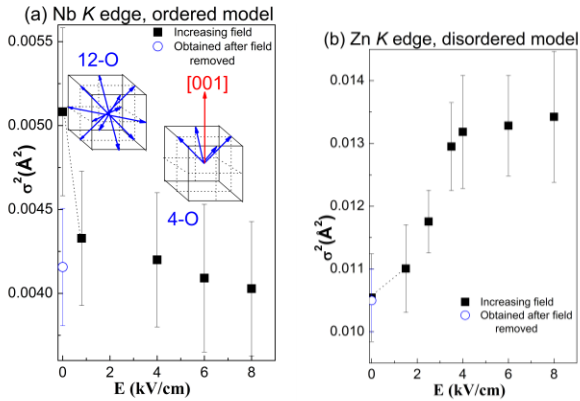

**Figure 8**. (a) Electric field dependence of the extracted $\sigma^2$(Nb-O). Disordered (12-O) and ordered polarized dipoles (4-O) are shown in the insets[32]. (b) Electric field dependence of $\sigma^2$(Zn-O). Filled squares correspond to points obtained on increasing field and the open circle corresponds to the point obtained after the field removed.

Additionally, $\sigma^2$(Nb-O) is found to smoothly decrease with increasing electric field and it retains the high field value when the electric field is removed. It is important to note that the refined Nb-O distances ($R_1$ and $R_2$) are quite stable over the entire electric field range, i.e. ($R_1$ = 1.94 (1) Å), $R_2$ = 2.15 (1) Å), which shows that the orthorhombic distortion remains relatively constant as a function of electric field. This result using the orthorhombic ordered model can be interpreted based on a general macroscopic view of the ferroelectric first hysteresis loop[33]. When the crystal is unpoled, it is consistent with a zero-macroscopic polarisation. All the dipoles of the orthorhombically distorted NbO$_6$ octahedra are disordered along the twelve equivalent <011> directions (12-O)[32] in Figure 8a, resulting in a global μNb = 0. When the sample is poled under electric field, all the dipoles are aligned along the 4 equivalent <101> direction close to the orientation of the electric field, (4-O) in Figure 8a. The decrease in $\sigma^2$(Nb-O) is therefore linked with a decrease in the width of the distribution of the orientations of the long Nb-O bonds due to the ordering of the dipoles with electric field. This causes corresponding decrease of the disorder which is consistent with the observed decrease of $\sigma^2$(Nb-O). When the field is removed, $\sigma^2$(Nb-O) remains unchanged because the dipoles stay aligned (the sample is still poled) [34]. $\sigma^2$(R1)= $\sigma^2$(R2) is justified by the fact that the static contribution dominates. This static contribution arises from a statistic distribution of μNb dipolar domain orientations.

Figure 8b shows the electric field dependence of the experimental $\sigma^2$(Zn-O) using a disordered cubic model. A noticeable increase of $\sigma^2$(Zn-O) can be observed with increasing $E$ followed by a flat slope about 0.0131 Å$^2$ above 3 kV/cm. It is interesting to note that electric field value is close to that of the coercive field ($E_C$). Additionally when the electric field is removed, $\sigma^2$(Zn-O) comes back to its initial value (0.0105 Å$^2$) while the refined Zn-O average distance remains almost constant, i.e. R = 2.066 (5) Å. In this study, as the thermal contribution is constant, changes in $\sigma^2$(Zn-O) with electric field corresponds to a structural contribution (i.e. polar displacement), which cannot be detected within the experimental accuracy of the EXAFS measurements with the used limited momentum transfer range, but can clearly be seen in the increase in the static contribution, which is significantly greater than the esd. As already discussed, adding a distortion with 2 distances $R_1$ and $R_2$ could be useful above 4 kV/cm although the estimated standard deviation was found to be of the order of the proposed distortion due to the limited $k$-range of 2 - 10 Å$^{-1}$. This indicates that this increase in $\sigma^2$(Zn-O) as a function of the electric field using a symmetric distribution of bond distances can be linked with a small, electric-field-induced polar shift. For E ≥ 4 kV/cm, this polar structural distortion reaches its critical limit which could be defined as the saturation of this polar displacement. Additionally, this small polar shift is found to be reversible, Figure 8b. Based on the results obtained by EXAFS at the Nb and Zn $K$ edges, we can propose a simplified scheme of the $B$-site behaviour under electric field in the PZN poled sample. As the electric field dependence of Zn-O and Nb-O bond lengths are found to be within the estimated standard deviation, the critical parameter appears to be the DWF, which in this study accounts for the static distribution of the corresponding Zn-O and Nb-O bonds. In the $AB'B''O_3$ structure of PZN, small Nb atoms are at the origin of strong dipoles along the <011>$_{pc}$ (with <011> pseudocubic orientations) directions. When the electric field is applied along <001>$_{pc}$ direction of a poled sample, Figure 9, the electric field will induce a small polar shift to the larger Zn atoms which will create a favourable energetic situation for the alignment of the strong dipoles bonds along <011>$_{pc}$ close to the orientation of the electric field. As this small polar shift can be obtained with a modest electric field, the corresponding $E_C$ to align the μNb dipoles is relatively low, which may account for the large piezoelectric effect in PZN. Additionally, this small polar shift of Zn is reversible, which means that it can change its direction when the applied electric field is reversed; such a reversible easy switching can clearly explain the flipping of the spontaneous ferroelectric μNb polarisation, Figure 9. In this simplified scheme of the $B$-site behaviour, the Zn-polar shift could be qualitatively viewed as the difference between the spontaneous polarisation at saturation and the remnant polarisation: $P_{S(saturation)}$-$P_{S(remnant)}$.

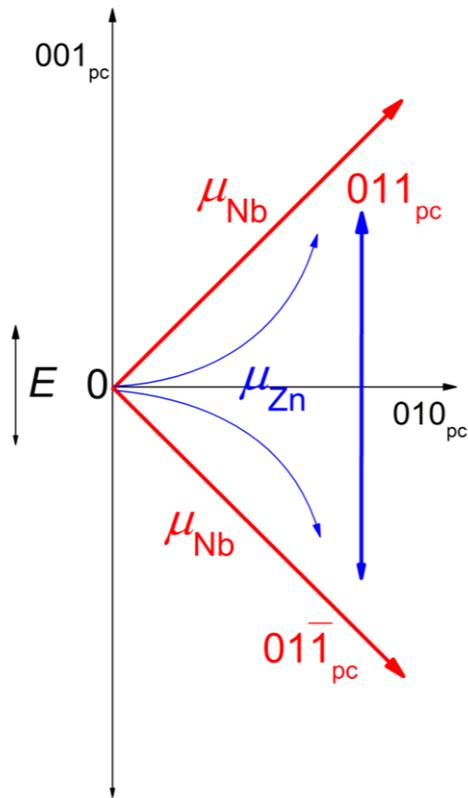

**Figure 9**. Simplified scheme of the *B*-site behaviour under electric field in the PZN poled sample.

## Conclusions

To conclude an "in situ" EXAFS study as a function of the applied electric field was undertaken on giant piezoelectric PZN single crystals at both the Nb and Zn *K* edges to understand the atomic scale behavior of the *B*-site atoms under the conditions of use in technological applications. We find that the small Nb atoms are at the origin of highly polarized Nb-O bonds, which are aligned in the poled state.

Alignment of the dipoles can be induced by a small polar shift of the Zn atoms under electric field, which plays an important role in the small $E_C$ of PZN and thus explains the giant piezoelectric response. As this small polar shift is found to be reversible as a function of the electric field, it therefore plays a major role in the polarization flipping when the direction of the electric field is reversed. Finally, such a mechanism could also be generalized in other ferroelectric materials which also present high or giant piezoelectric effect, the $PbMg_{1/3}Nb_{2/3}O_3$ (PMN), $(Pb_{1-y}La_y)Zr_{1-x}Ti_xO_3$ (P(L)ZT), $BiScO_3$-$PbTiO_3$ solid solutions[35]...

## ASSOCIATED CONTENT

**Supporting Information**

## AUTHOR INFORMATION

**Corresponding Author**

Jerome.Rouquette@univ-montp2.fr

**Acknowledgments**

The authors would like to thank the beamline staff at BM29 at the ESRF for their assistance and Alexandre Lebon who synthetized the single crystals of PZN during its PhD thesis.

**Author Contributions**

(Word Style "Section_Content").    ‡These authors contributed equally. (match statement to author names with a symbol, if applicable)

**Notes**

The authors declare no competing financial interests.
Any additional relevant notes should be placed here.

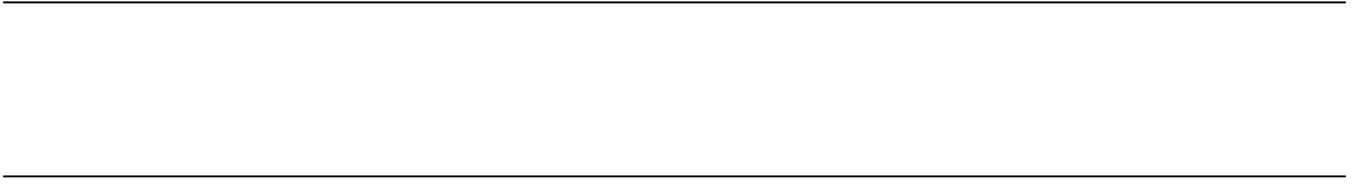
Insert Table of Contents artwork here